\documentclass[11pt]{article}
\usepackage{amsbsy,amsfonts,amsmath,amssymb,amsthm,euscript,enumerate}
\usepackage{graphicx,color}
\usepackage{bbm}
\usepackage{url}
\usepackage{hyperref}
\hypersetup{colorlinks,citecolor=blue,linkcolor=blue,urlcolor=black}
\newcommand{\indic}{\mathbbm{1}}

\allowdisplaybreaks[1]

\theoremstyle{plain}
\numberwithin{equation}{section}
\newtheorem{thm}{Theorem}[section]

\newtheorem{lmm}[thm]{Lemma}
\newtheorem{prp}[thm]{Proposition}

\newcommand{\cC}{\mathcal{C}}
\newcommand{\cZ}{\mathcal{Z}}
\newcommand{\dpst}{\displaystyle}
\newcommand{\Exp}[1]{\langle #1 \rangle}
\newcommand{\ind}[1]{\indic_{\{#1\}}}
\newcommand{\lbeq}[1]{\label{eq:#1}}
\newcommand{\mE}{{\mathbb E}}
\newcommand{\mP}{{\mathbb P}}

\newcommand{\nn}{\nonumber}

\newcommand{\pc}{p_\mathrm{c}}
\newcommand{\Proof}[1]{\paragraph{\it #1}}
\newcommand{\QED}{\hspace*{\fill}\rule{7pt}{7pt}\smallskip}
\newcommand{\refeq}[1]{(\ref{eq:#1})}
\newcommand{\sss}{\scriptscriptstyle}
\newcommand{\Tc}{T_{\rm c}}
\newcommand{\veee}[1]{\|\hskip-1pt|#1\|\hskip-1pt|}
\newcommand{\Veee}[1]{\big\|\hskip-1pt\big|#1\big\|\hskip-1pt\big|}
\newcommand{\vep}{\varepsilon}
\newcommand{\vno}{\varnothing}
\newcommand{\vphi}{\varphi}
\newcommand{\Z}{\mathbb{Z}}
\newcommand{\Zd}{\Z^d}

\newcommand{\bl}{\boldsymbol{l}}
\newcommand{\bm}{\boldsymbol{m}}
\newcommand{\bn}{\boldsymbol{n}}

\newcommand{\bsigma}{\boldsymbol{\sigma}}

\newcommand{\cn}[2]
{\dpst\mathop{\longleftrightarrow}_{\raisebox{3pt}{$\sss #1$}}^{#2}}
\newcommand{\ncn}[2]
{\dpst\mathop{\,\,\,\not\!\!\!\longleftrightarrow}_{\raisebox{3pt}{$\sss #1$}}^{#2}}

\title{Mean-field bound on the 1-arm exponent for Ising ferromagnets
in high dimensions}
\author{
Satoshi Handa\footnote{Graduate School of Mathematics, Hokkaido University,
Japan. {\tt handa@math.sci.hokudai.ac.jp}},~
Markus Heydenreich\footnote{Mathematisches Institut,
Ludwig-Maximilians-Universit\"at M\"unchen, Germany.
{\tt m.heydenreich@lmu.de}},~
Akira Sakai\footnote{Faculty of Science, Hokkaido University,
Japan. {\tt sakai@math.sci.hokudai.ac.jp}, 
\url{https://orcid.org/0000-0003-0943-7842}}
}

\begin{document}
\maketitle

\begin{abstract}
The 1-arm exponent $\rho$ for the ferromagnetic Ising model on $\Zd$ is the
critical exponent that describes how fast the critical 1-spin expectation at
the center of the ball of radius $r$ surrounded by plus spins decays in powers
of $r$.  Suppose that the spin-spin coupling $J$ is translation-invariant,
$\Zd$-symmetric and finite-range.  Using the random-current representation
and assuming the anomalous dimension $\eta=0$, we show that the optimal
mean-field bound $\rho\le1$ holds for all dimensions $d>4$.
This significantly improves a bound previously obtained by a hyperscaling
inequality.
\end{abstract}

\begin{center}
\large\emph{We dedicate this work to Chuck Newman on the occasion of his 70th birthday.}
\end{center}

\begin{minipage}{15cm}\small
\textbf{A personal note.}
I (AS) met Chuck for the first time when I visited New York for a month in
summer 2004.  I knew him before, not in person, but for his influential papers
on phase transitions and critical behavior of percolation and the Ising model.
So, naturally, I felt awful to speak with him (partly because of my poor
language skills) and to give a presentation on my work back then.
However, he was friendly and positive about my presentation, which he may
no longer remember.  Since then, I became a big fan of his.  There must be
many researchers like me who were greatly encouraged by Chuck, and I am sure
there will be many more.

\qquad The topic of my presentation back in summer 2004 was about the critical 
exponent $\rho$ for the percolation 1-arm probability in high dimensions 
\cite{s04}.  The 1-arm probability is the probability that the center of the 
ball of radius $r$ is connected to its surface by a path of occupied bonds.  
Compared with most of the other critical exponents, such as $\beta,\gamma,\eta$ 
and $\delta$, the 1-arm exponent $\rho$ is harder to investigate, not because 
we still do not know the continuity of the critical percolation probability (= 
the $r\uparrow\infty$ limit of the critical 1-arm probability), but because 
$\rho$ is associated with a finite-volume quantity and therefore deals
with boundary effects. Even in high dimensions, it is difficult
to identify the mean-field value of $\rho$.  However, by the second-moment
method \cite{s04}, it is rather easy to show the one-sided inequality
$\rho\le2$, if $\rho$ exists and $\eta=0$.  The latter assumption is known
to hold in high dimensions, thanks to the lace-expansion results
\cite{h08,hhs03}.  Then, in \cite{kn11}, Kozma and Nachmias finally proved
the equality $\rho=2$ by assuming $\eta=0$ and using a sophisticated
inductive argument.
\end{minipage}

\section{Introduction}
We consider the ferromagnetic Ising model at its critical temperature $T=T_c$,
and study the 1-spin expectation $\Exp{\sigma_o}_r^+$ at the center of a ball
of radius $r$ surrounded by plus spins.
The decreasing limit of $\Exp{\sigma_o}_r^+$ as $r\uparrow\infty$ is the
spontaneous magnetization. Recently, Aizenman, Duminil-Copin and Sidoravicius
\cite{ads15} showed that, if the spin-spin coupling satisfies a strong
symmetry condition called reflection-positivity, then the spontaneous
magnetization is a continuous function of temperature in all dimensions $d>2$,
in particular $\lim_{r\uparrow\infty}\Exp{\sigma_o}_r^+=0$ at criticality.
The present paper gives quantitative bounds on the rate of convergence.
The nearest-neighbor model is an example that satisfies reflection-positivity.
Also, its spontaneous magnetization on $\mathbb{Z}^2$ is known to be zero
at criticality \cite{y52}.  However, in general, finite-range models do not
satisfy reflection-positivity, and therefore we cannot automatically justify
continuity of the spontaneous magnetization for, e.g., the
next-nearest-neighbor model.  Fortunately, by using the
lace expansion \cite{cs15,s07}, we can avoid assuming reflection-positivity
to ensure $\eta=0$ (as well as $\beta=1/2$, $\gamma=1$, $\delta=3$) and
$\lim_{r\uparrow\infty}\Exp{\sigma_o}_r^+=0$ at criticality in dimensions
$d>4$ if the support of $J$ is large enough.

In this paper, we prove that it does not decay very fast whenever $d>4$;
in this case we prove $\Exp{\sigma_o}_r^+\ge r^{-1+o(1)}$.
The proof relies on the random-current representation,
which is a sophisticated version of the high-temperature expansion.
It was initiated in \cite{ghs70} to show the GHS inequality.
Then, in 1980's, Aizenman revived it to show that the bubble condition
(i.e., square-summability of the critical 2-spin expectation)
is a sufficient condition for the mean-field behavior \cite{a82,abf87,af86}.
It is also used in \cite{ads15,s07,s15} to obtain many useful results for the
Ising and $\vphi^4$ models.  In combination with the second-moment method,
we prove a correlation inequality that involves $\Exp{\sigma_o}_r^+$ and
free-boundary 2-spin expectations.  Then, by using this correlation inequality,
we derive the desired result.

First, we provide the precise definition of the model.

\subsection{The model}
First we define the Ising model on $V_R$, which is the $d$-dimensional ball
of radius $R>0$:
\begin{align}
V_R=\{v\in\Zd:|v|\le R\}.
\end{align}
It is convenient to use the Euclidean distance $|\cdot|$ here,
but our results hold for any norm on the lattice $\Zd$.
We define the Hamiltonian for a spin configuration
$\bsigma\equiv\{\sigma_v\}_{v\in V_R}\in\{\pm1\}^{V_R}$ as
\begin{align}
H_{r,R}^h(\bsigma)=-\sum_{\{u,v\}\subset V_R}J_{u,v}\sigma_u\sigma_v
 -h\sum_{v\in\partial V_r}\sigma_v,
\end{align}
where $J_{u,v}\ge0$ is a translation-invariant, $\Zd$-symmetric and
finite-range coupling, $h$ is the strength of the external magnetic field, 
and $\partial V_r$ ($r<R$) is the boundary of $V_r$:
\begin{align}
\partial V_r=\{v\in V_R\setminus V_r:\exists u\in V_r\text{ such that }
 J_{u,v}>0\}.
\end{align}
We note that it is crucial to impose the external magnetic field only on
$\partial V_r$.  Due to this slightly unusual setup, we will eventually be
able to derive an essential correlation inequality that differs from the one
for percolation.

The thermal expectation of
a function $f$ on spin configurations at the critical temperature $\Tc$ is
given by
\begin{align}
\Exp{f}_{r,R}^h=\frac1{2^{|V_R|}}\sum_{\bsigma\in\{\pm1\}^{V_R}}f(\bsigma)\,
 \frac{e^{-H_{r,R}^h(\bsigma)/\Tc}}{Z_{r,R}^h},&&
Z_{r,R}^h=\frac1{2^{|V_R|}}\sum_{\bsigma\in\{\pm1\}^{V_R}}e^{-H_{r,R}^h
 (\bsigma)/\Tc}.
\end{align}
The major quantities to be investigated are the 1-spin and 2-spin expectations.
Since they are increasing in $h$ by Griffiths' inequality \cite{g70}, we simply
denote their limits by
\begin{align}
\Exp{\sigma_x}_r^+&=\lim_{h\uparrow\infty}\Exp{\sigma_x}_{r,R}^h
 \qquad[x\in V_r\cup\partial V_r],\\
\Exp{\sigma_x\sigma_y}_R&=\lim_{h\downarrow0}\Exp{\sigma_x\sigma_y}_{r,R}^h
 \qquad[x,y\in V_R].
\end{align}
Since $\Exp{\sigma_x\sigma_y}_R$ is also increasing in $R$ by Griffiths'
inequality, we denote its limit by
\begin{align}
\Exp{\sigma_x\sigma_y}=\lim_{R\uparrow\infty}\Exp{\sigma_x\sigma_y}_R.
\end{align}

In the following statement (as well as later in the proofs) we use the
notation $f\asymp g$ to mean that the ratio $f/g$ is bounded away from zero
and infinity (in the prescribed limit).  One assumption that we shall make
throughout is the mean-field decay for the critical 2-spin expectation 
(or often called two-point function)
\begin{align}\lbeq{twopt}
\Exp{\sigma_o\sigma_x}\asymp|x|^{2-d}\quad\text{as}~|x|\uparrow\infty.
\end{align}
A sharp asymptotic expression that implies \refeq{twopt} is proven by the
lace expansion for a fairly general class of $J$, whenever the support of
$J$ is sufficiently large \cite{s07}.  We note that reflection positivity has
not succeeded in providing the above two-sided $x$-space bound; only one
exception is the nearest-neighbor model, for which a one-sided $x$-space
bound is proven \cite{s82}.  In dimensions $d<4$, the exponent on the
right-hand side may change.  An exact solution for $d=2$ was identified by Wu
et al.~\cite{wmtb76}, which implies $\Exp{\sigma_o\sigma_x}\asymp|x|^{-1/4}$
as $|x|\uparrow\infty$.

\subsection{The main result}
We are investigating the 1-arm exponent for the Ising model at criticality,
informally described as $\Exp{\sigma_o}_r^+\approx r^{-\rho}$ as
$r\uparrow\infty$.  In order to make the symbol $\approx$ precise,
we give the formal definition
\begin{align}\lbeq{rho-def}
	\rho=-\liminf_{r\to\infty}\frac{\log\Exp{\sigma_o}_r^+}{\log r}.
\end{align}
A more conventional way of defining $\rho$ is by letting
$\Exp{\sigma_o}_r^+\asymp r^{-\rho}$ as $r\uparrow\infty$,
which was used to define the percolation 1-arm exponent \cite{kn11,s04}.
However, the latter definition does not necessarily guarantee
the existence of $\rho$.  To avoid this existence issue,
we adopt the former definition \refeq{rho-def}.

Our main result is the one-sided bound $\rho\le1$ in the mean-field regime,
i.e., when $d>4$ and \refeq{twopt} holds.  Folklore of statistical physics
predicts that \refeq{rho-def} is actually a limit. The use of limit inferior
is somewhat arbitrary ($\limsup$ would be also possible), but this choice
gives the strongest result.

\begin{thm}\label{thm:main}
For the ferromagnetic Ising model on $\Zd$, $d>4$, defined by a
translation-invariant, $\Zd$-symmetric and finite-range spin-spin coupling
satisfying \refeq{twopt},
\begin{align}\lbeq{main}
\liminf_{r\to\infty}r^{1+\vep}\Exp{\sigma_o}_r^+ =\infty
\end{align}
whenever $\vep>0$.  Consequently, the critical exponent $\rho$ defined in
\refeq{rho-def} satisfies $\rho\le1$.
\end{thm}

Tasaki \cite{t87} proved that
$\Exp{\sigma_o}_{|x|/3}^+\ge\sqrt{\Exp{\sigma_o\sigma_x}}$ holds at any
temperature for sufficiently large $|x|$ (so that $|x|/3$ is larger than the
range of the spin-spin coupling).  In dimensions $d>4$, this implies the
hyperscaling inequality $\rho\le(d-2)/2$, and our bound in \refeq{main}
improves on Tasaki's result.

It is a challenge now to prove
\begin{align}\lbeq{upperbd}
\limsup_{r\to\infty}r^{1-\vep}\Exp{\sigma_o}_r^+ =0
\end{align}
for any $\vep>0$, which implies readily (together with our theorem) that
\refeq{rho-def} is actually a limit and $\rho=1$.

Our proof of $\rho\le1$ uses \refeq{twopt}, which requires $d>4$ (and the
support of $J$ to be large), even though the result is believed to be true for
all dimensions $d\ge2$.  The aforementioned correlation inequality
$\Exp{\sigma_o}_{|x|/3}^+\ge\sqrt{\Exp{\sigma_o\sigma_x}}$ combined with
the exact solution for $d=2$ \cite{wmtb76} and numerical predictions for
$d=3,4$ supports this belief.  This is why we call $\rho\le1$ the optimal
mean-field bound.

Another key ingredient for the proof of $\rho\le1$ is the random-current
representation, which provides a translation between spin correlations and
percolation-like connectivity events.  Then, by applying the second-moment
method to the connectivity events as explained below for percolation,
we can derive a crucial correlation inequality (cf. \refeq{correlation})
that relates 1-spin and 2-spin expectations.  To explain what the
second-moment method is and to compare the resulting correlation
inequalities for the two models, we spend the next subsection to explain
the derivation of the mean-field bound on the percolation 1-arm exponent,
i.e., $\rho\le2$ for $d>6$.

\subsection{Derivation of the mean-field bound for percolation}
We consider the following bond percolation on $\Zd$.  Each bond
$\{u,v\}\subset\Zd$ is either occupied or vacant with probability $pJ_{u,v}$
or $1-pJ_{u,v}$, independently of the other bonds, where $p\ge0$ is the
percolation parameter.  The 2-point function $G_p(x,y)$ is the probability
that $x$ is connected to $y$ by a path of occupied bonds ($G_p(x,x)=1$
by convention).  It is well-known that, for any $d\ge2$, there is a
nontrivial critical point $\pc$ such that the susceptibility $\sum_xG_p(o,x)$
is finite if and only if $p<\pc$ \cite{an84}.
The 1-arm probability $\theta_r$, which is the probability that the center
of the ball of radius $r$ is connected to its surface by a path of occupied
bonds, also exhibits a phase transition at $\pc$ \cite{ab87}:
$\theta(p)\equiv\lim_{r\uparrow\infty}\theta_r=0$ if $p<\pc$ and
$\theta(p)>0$ if $p>\pc$.  Although the continuity $\theta(\pc)=0$
has not yet been proven in full generality, it is shown by the lace expansion
\cite{fh16,hs90} that, if $d>6$ and the support of $J$ is sufficiently large,
then $\theta(\pc)=0$ and $G_{\pc}(o,x)\asymp|x|^{2-d}$ as
$|x|\uparrow\infty$.  This Newtonian behavior of $G_{\pc}$ is believed
not to hold in lower dimensions.

Fix $p=\pc$ and define the percolation 1-arm exponent $\rho$ by letting
$\theta_r\asymp r^{-\rho}$ as $r\uparrow\infty$.  Since
$\theta_{|x|/3}\ge\sqrt{G_p(o,x)}$ if $|x|\gg1$ \cite{t87},
it is known that the same hyperscaling inequality $\rho\le(d-2)/2$ holds
for all dimensions $d>6$.  In \cite{s04}, we were able to improve this to
the optimal mean-field bound $\rho\le2$ for $d>6$ by using the
second-moment method, which we explain now.  Let $X_r$ be the
random number of vertices on $\partial V_r$ that are connected to
the origin $o$.  We note that $X_r$ can be positive only when $o$ is
connected to $\partial V_r$.  Then, by the Schwarz inequality,
\begin{align}
\mE_p[X_r]^2=\mE_p\Big[X_r\ind{o\text{ is connected to }\partial V_r}\Big]^2
 \le\underbrace{\mE_p\Big[\ind{o\text{ is connected to }\partial V_r}
 \Big]}_{=\theta_r}\mE_p[X_r^2],
\end{align}
which implies $\theta_r\ge\mE_p[X_r]^2/\mE_p[X_r^2]$.  Notice that
$\mE_p[X_r]=\sum_{x\in\partial V_r}G_p(o,x)$ and that, by the tree-graph
inequality \cite{an84},
\begin{align}
\mE_p[X_r^2]=\sum_{x,y\in\partial V_r}\mP_p(o\text{ is connected to }x,y)
 \le\sum_{\substack{u\in\Zd\\ x,y\in\partial V_r}}G_p(o,u)\,G_p(u,x)\,G_p(u,y).
\end{align}
As a result, we arrive at the correlation inequality
\begin{align}\lbeq{perc-correlation}
\theta_r\ge\frac{\dpst\bigg(\sum_{x\in\partial V_r}G_p(o,x)\bigg)^2}{\dpst
 \sum_{\substack{u\in\Zd\\ x,y\in\partial V_r}}G_p(o,u)\,G_p(u,x)\,G_p(u,y)}.
\end{align}
Using $G_{\pc}(x,y)\asymp\veee{x-y}^{2-d}$, where $\veee\cdot=|\cdot|\vee1$ is
to avoid singularity around zero, we can readily show that the right-hand side
of the above inequality is bounded from below by a multiple of $r^{-2}$,
resulting in $\rho\le2$ for $d>6$.

In order to prove the opposite inequality $\rho\ge2$ for $d>6$ to conclude the
equality, Kozma and Nachmias \cite{kn11} use another correlation inequality
that involves not only $\theta_r$ and $G_p$ but also the mean-field
cluster-size distribution.  The Ising cluster-size distribution under the
random-current representation is not available yet, and we are currently
heading in that direction.

\subsection{Further discussion}
\begin{enumerate}
\item
\emph{On trees.}
The Ising model on trees, also known as the Ising model on the Bethe lattice,
is rigorously studied since the 1970's \cite{kt74,p74}.   In contrast to
amenable graphs, the phase transition on trees can appear even when
there is a non-zero homogeneous random field, cf. \cite{js99}.
One line of research considers critical-field Ising models under
the influence of an inhomogeneous external field, for which we refer to
the discussion in \cite{bee16}.
The absence of loops in the underlying graph makes it easier to analyze,
and a number of critical exponents are known to take on their mean-field
values, see \cite[Section~4.2]{s01}.

For the Ising model on a regular tree, it is shown \cite{hk18} that
\begin{equation}\lbeq{rho-tree}
	\Exp{\sigma_o}_r^+\asymp r^{-1/2}\quad\text{as}~r\uparrow\infty,
\end{equation}
where, instead of the ball $V_r$, we are using the subtree of depth $r$ from 
the root (with the plus-boundary condition).  This proves $\rho=1/2$ on trees.  
The discrepancy to the high-dimensional setting can be resolved by adjusting 
the notion of distance in the tree, that is, one should rather work with the 
metric $\mathrm{dist}(o,x):=\sqrt{\mathrm{depth}(x)}$ incorporating spatial
effects when embedding the tree into the lattice $\Zd$.
With this notion of distance, we get the mean-field value $\rho=1$.
The same situation occurs for percolation, where we refer to
\cite{g99,hh17} and  for a discussion of this issue.
\item
\emph{Long-range models.}
In our result, we assumed that the spin-spin coupling $J$ is finite-range,
that is, there is an $M>0$ such that $J_{o,x}=0$ whenever $|x|>M$.
We believe that $\rho=1$ is true even for infinite-range couplings with
sufficiently fast decaying tails, although the boundary $\partial V_r$
on which the external magnetic field is imposed under the current setting
is no longer bounded.  The situation may change when we consider couplings
with regularly-varying tails, and we focus now on the situation when
\begin{equation} \lbeq{longrangecoupling}
	 J_{o,x}\asymp |x|^{-d-\alpha} \quad\text{as}~|x|\uparrow\infty,
\end{equation}
for some $\alpha>0$.  In our earlier work \cite{cs15}, we show that,
under a suitable spread-out condition, the critical 2-spin expectation scales as
\begin{align}
\Exp{\sigma_o\sigma_x}\asymp|x|^{\alpha\wedge2-d}\quad
 \text{as}~|x|\uparrow\infty,
\end{align}
in contrast to \refeq{twopt}.  In particular, there is a crossover at
$\alpha=2$ between a ``finite-range regime" and a ``long-range regime".

For the critical exponent $\rho$, it is tempting to believe that this crossover
happens for $\alpha=4$.  The reason for this is again a comparable result
for percolation: Hulshof \cite{h15} proved that, if
$G_{\pc}(o,x)\asymp|x|^{\alpha\wedge2-d}$ as $|x|\uparrow\infty$,
then the critical 1-arm probability scales as
$\theta_r\asymp r^{-(\alpha\wedge4)/2}$ as $r\uparrow\infty$.
In view of Hulshof's result, it is plausible that, for the long-range Ising
model with couplings like in \refeq{longrangecoupling}, it is the case
that $\rho=(\alpha\wedge4)/4$.
\item
\emph{The (1-component) $\vphi^4$ model.}
This spin model is considered to be in the same universality class as Ising
ferromagnets \cite{a82}.  It can be constructed as an $N\uparrow\infty$ limit
of a properly coupled $N$ ferromagnetic Ising systems \cite{sg73},
and therefore we can apply the random-current representation for the Ising
model.  By virtue of this representation, we can use the lace expansion to
show that the critical 2-spin expectation satisfies \refeq{twopt} for a large
class of short-range models \cite{s15}.  It is natural to be interested in
the critical 1-spin expectation similar to $\Exp{\sigma_o}_r^+$ for the
Ising model.  However, since the $\vphi^4$ spin is an unbounded variable,
we cannot simply take $h\uparrow\infty$ to define the 1-spin expectation
under the ``plus-boundary" condition.  Once it is defined appropriately,
we believe that its 1-arm exponent also satisfies the mean-field bound
$\rho\le1$ for $d>4$.
\end{enumerate}

From the next section, we begin the proof of the main theorem.
In Section~\ref{ss:RCrep}, we introduce notation and definitions
associated with the random-current representation.
In Section~\ref{ss:correlation}, we use the random-current representation
and the second-moment method to derive a key correlation inequality.
Finally, in Section~\ref{ss:proof}, we use the obtained correlation
inequality and \refeq{twopt} to conclude that $\rho\le1$ for $d>4$.

\section{Proof of the results}
\subsection{The random-current representation}\label{ss:RCrep}
\begin{figure}[t]
\begin{align*}
\Exp{\sigma_o}_{r,R}^h=\frac{\includegraphics[scale=0.3]{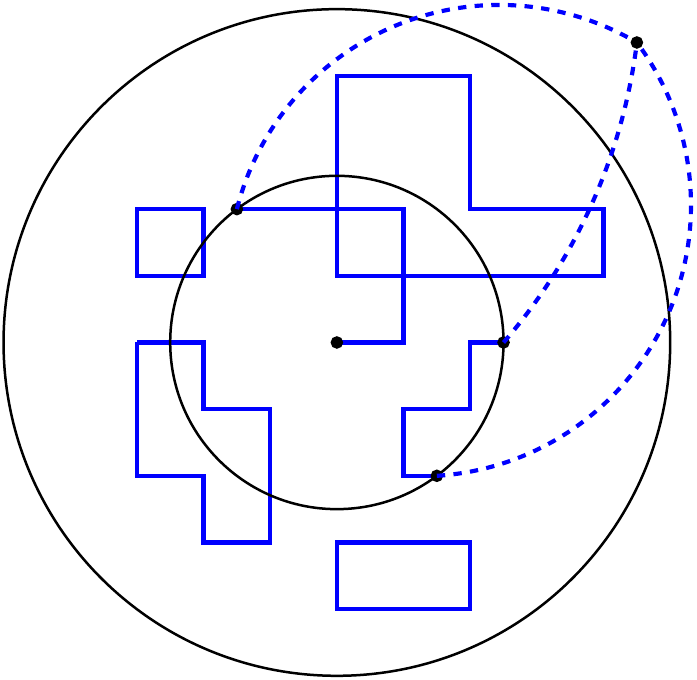}}
 {\includegraphics[scale=0.3]{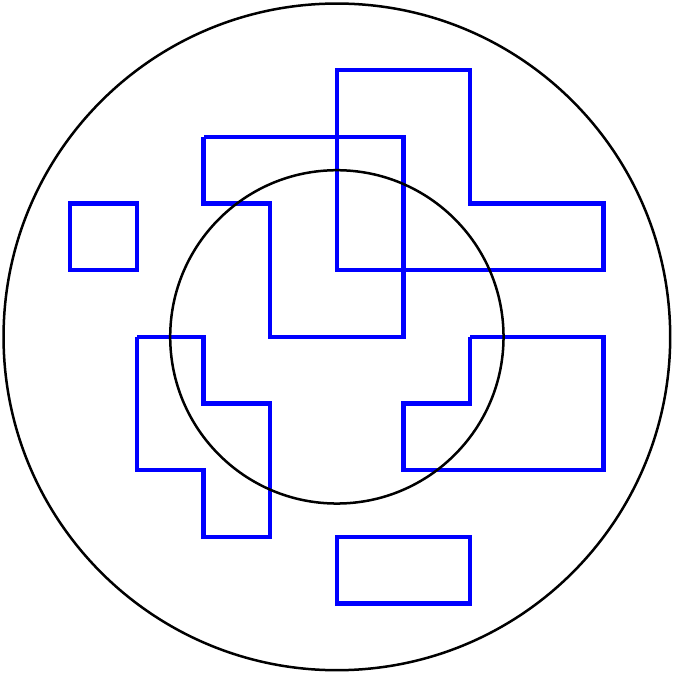}},&&
\Exp{\sigma_o\sigma_x}_R=\frac{\includegraphics[scale=0.3]{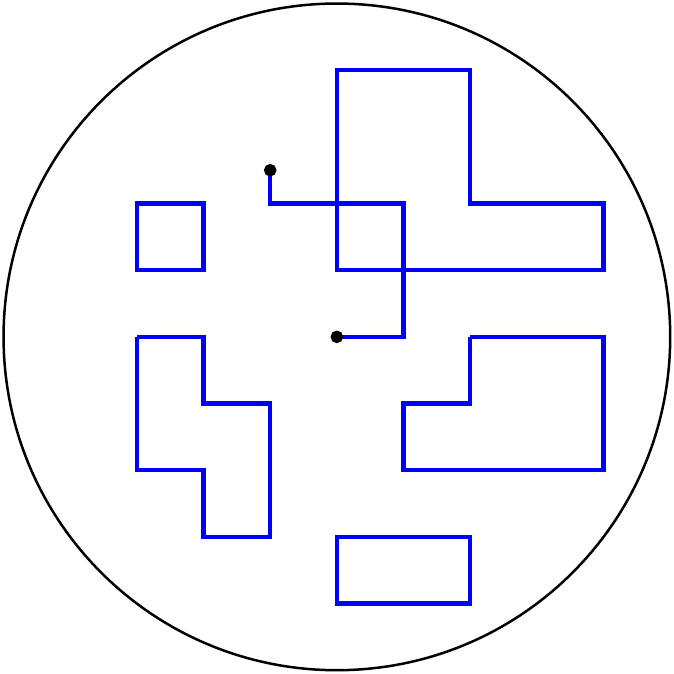}}
 {\includegraphics[scale=0.3]{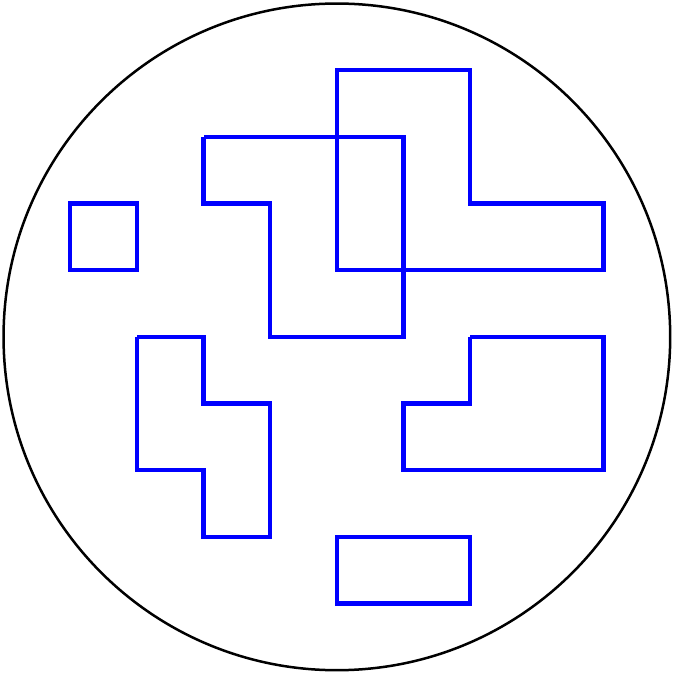}}.
\end{align*}
\caption{\label{fig:RCrep}The random-current representation for
$\Exp{\sigma_o}_{r,R}^h$ and $\Exp{\sigma_o\sigma_x}_R$.  The bonds with even
current are all omitted.  The vertex connected by dashed line segments is the
ghost site $g$.}
\end{figure}

A current configuration $\bn\equiv\{n_b\}$ is a set of nonnegative integers on
bonds $b\in B_R\equiv\big\{\{u,v\}\subset V_R:J_{u,v}>0\big\}$ or
$b\in G_r\equiv\big\{\{v,g\}:v\in\partial V_r\big\}$, where $g$ is an imaginary
ghost site.  Given a current configuration $\bn$, we define the source set
$\partial\bn$ as
\begin{align}
\partial\bn=\bigg\{v\in V_R\cup\{g\}:\sum_{b\ni v}n_b\text{ is odd}\bigg\},
\end{align}
and the weight functions $w_{r,R}^h(\bn)$ and $w_R(\bn)$ as
\begin{align}
w_{r,R}^h(\bn)=\prod_{b\in B_R}\frac{(J_b/\Tc)^{n_b}}{n_b!}\prod_{b'\in G_r}
 \frac{(h/\Tc)^{n_{b'}}}{n_{b'}!},&&
w_R(\bn)=w_{r,R}^0(\bn).
\end{align}
Then, we obtain the following random-current representation
(cf. Figure~\ref{fig:RCrep}):
\begin{align}
Z_{r,R}^h=\sum_{\partial\bn=\vno}w_{r,R}^h(\bn),&&
Z_R=\sum_{\partial\bn=\vno}w_R(\bn),
\end{align}
and for $x,y\in V_R$,
\begin{align}
\Exp{\sigma_x}_{r,R}^h=\sum_{\partial\bn=\{x,g\}}\frac{w_{r,R}^h(\bn)}
 {Z_{r,R}^h},&&
\Exp{\sigma_x\sigma_y}_R=\sum_{\partial\bn=\{x\}\triangle\{y\}}\frac{w_R(\bn)}
 {Z_R}.
\end{align}
Given a current configuration $\bn=\{n_b\}$, we say that $x$ is
$\bn$-connected to $y$,
denoted $x\cn{\bn}{}y$
if either $x=y\in V_R\cup\{g\}$
or there is a path from $x$ to $y$ consisting of bonds $b\in B_R\cup G_r$
with $n_b>0$. For $A\subset V_R\cup\{g\}$, we also say that $x$ is
$\bn$-connected to $y$ in $A$, denoted $x\cn{\bn}{}y$ in  $A$,
if either $x=y\in A$ or there is a path from $x$ to $y$ consisting of
bonds $b\subset A$ with $n_b>0$.

Given a subset $A\subset V_R$, we define
\begin{align}
W_A(\bm)=\prod_{b\in A}\frac{(J_b/\Tc)^{m_b}}{m_b!},&&
\cZ_A=\sum_{\partial\bm=\vno}W_A(\bm).
\end{align}

The most important feature of the random-current representation is the
so-called source-switching lemma (e.g., \cite[Lemma~2.3]{s07}).
We state the version we use the most in this paper as below. This is an immediate consequence from the source-switching lemma.

\begin{lmm}[{Consequence from the source-switching lemma, \cite{s07}}]\label{lmm:SSL}
For any subsets $A\subset V_R$ and $B\subset V_R\cup\{g\}$, any $x,y\in V_R$ 
and any function $f$ on current configurations,
\begin{align}\lbeq{SSL}
\sum_{\substack{\partial\bn=B\\ \partial\bm=\vno}}w_{r,R}^h(\bn)\,W_A(\bm)\indic\{x\cn{\bn+\bm}{}y~\mathrm{in}~A\}\,f(\bn+\bm)
=\sum_{\substack{\partial\bn=B\bigtriangleup\{x\}\bigtriangleup\{y\}\\ \partial\bm=\{x\}\bigtriangleup\{y\}}}w_{r,R}^h(\bn)\,W_A(\bm)\,f(\bn+\bm).
\end{align}
\end{lmm}
For a proof, we refer to \cite[Lemma~2.3]{s07}.

\subsection{A correlation inequality}\label{ss:correlation}
The main technical vehicle in the proof of Theorem \ref{thm:main} is the
following  correlation inequality that relates $\Exp{\sigma_o}_r^+$ to
the sum of 2-spin expectations.

\begin{prp}\label{prp:2ndmom}
For the ferromagnetic Ising model,
\begin{align}\lbeq{correlation}
\Exp{\sigma_o}_r^+\ge\frac{\dpst\bigg(\sum_{x\in\partial V_r}\Exp{\sigma_o
 \sigma_x}\bigg)^2}{\dpst\sum_{x,y\in\partial V_r}\Exp{\sigma_o\sigma_x}
 \Exp{\sigma_x\sigma_y}+\sum_{\substack{u\in\Zd\\ x,y\in\partial V_r}}
 \Exp{\sigma_o\sigma_u}\Exp{\sigma_u\sigma_x}\Exp{\sigma_u\sigma_y}
 \Exp{\sigma_o}_{\mathrm{dist}(u,\partial V_r)}^+}.
\end{align}
\end{prp}

Compare this with the correlation inequality \refeq{perc-correlation} for
percolation.  The extra factor in the denominator of \refeq{correlation},
$\Exp{\sigma_o}_{\mathrm{dist}(u,\partial V_r)}^+$, will eventually be the
key to obtain the optimal mean-field bound on the Ising 1-arm exponent.

\Proof{Proof of Proposition~\ref{prp:2ndmom}.}
The proof is carried out in four steps.

\paragraph{\it Step 1: The second-moment method.}
Let
\begin{align}
X_r(\bn)=\sum_{x\in\partial V_r}\indic\{o\cn{\bn}{}x\text{ in }V_R\}.
\end{align}
Then, by the Schwarz inequality, we obtain
\begin{align}\lbeq{schwarz}
\sum_{\substack{\partial\bn=\{o,g\}\\ \partial\bm=\vno}}\frac{w_{r,R}^h(\bn)}
 {Z_{r,R}^h}\,\frac{w_R(\bm)}{Z_R}\,X_r(\bn+\bm)&\le\bigg(
 \underbrace{\sum_{\partial\bn=\{o,g\}}\frac{w_{r,R}^h(\bn)}{Z_{r,
 R}^h}}_{=\Exp{\sigma_o}_{r,R}^h}\,\underbrace{\sum_{\partial\bm=\vno}
 \frac{w_R(\bm)}{Z_R}}_{=1}\bigg)^{1/2}\nn\\
&\quad\times\bigg(\sum_{\substack{\partial\bn=\{o,g\}\\ \partial\bm=\vno}}
 \frac{w_{r,R}^h(\bn)}{Z_{r,R}^h}\,\frac{w_R(\bm)}{Z_R}\,X_r
 (\bn+\bm)^2\bigg)^{1/2}.
\end{align}
By Lemma~\ref{lmm:SSL}, we can rewrite the left-hand side as
\begin{align}\lbeq{SSL1}
\sum_{x\in\partial V_r}\sum_{\substack{\partial\bn=\{o,g\}\\ \partial\bm=\vno}}
 \frac{w_R(\bn)}{Z_{r,R}}\,\frac{w_R(\bm)}{Z_R}\,\indic\{o\cn{\bn
 +\bm}{}x\text{ in }V_R\}&=\sum_{x\in\partial V_r}\sum_{\substack{\partial\bn
 =\{x,g\}\\ \partial\bm=\{o,x\}}}\frac{w_{r,R}^h(\bn)}{Z_{r,R}^h}\,
 \frac{w_R(\bm)}{Z_R}\nn\\
&=\sum_{x\in\partial V_r}\Exp{\sigma_x}_{r,R}^h\,\Exp{\sigma_o\sigma_x}_R.
\end{align}
As a result, we obtain
\begin{align}\lbeq{1stbd}
\Exp{\sigma_o}_{r,R}^h\ge\frac{\dpst\bigg(\sum_{x\in\partial V_r}\Exp{\sigma_x}
 _{r,R}^h\,\Exp{\sigma_o\sigma_x}_R\bigg)^2}{\dpst\sum_{\substack{\partial\bn
 =\{o,g\}\\ \partial\bm=\vno}}\frac{w_{r,R}^h(\bn)}{Z_{r,R}^h}\,\frac{w_R
 (\bm)}{Z_R}\,X_r(\bn+\bm)^2}.
\end{align}

\paragraph{\it Step 2: Switching sources.}
Next, we investigate the denominator of the right-hand side of
\refeq{1stbd}, which equals
\begin{align}\lbeq{denom}
\sum_{x,y\in\partial V_r}\sum_{\substack{\partial\bn=\{o,g\}\\ \partial\bm
 =\vno}}\frac{w_{r,R}^h(\bn)}{Z_{r,R}^h}\,\frac{w_R(\bm)}{Z_R}\,
 \indic\{o\cn{\bn+\bm}{}x,y\text{ in }V_R\}.
\end{align}
The contribution from the summands $x=y$ may be rewritten as in \refeq{SSL1}.
Similarly, for the case of $x\ne y$, we use Lemma~\ref{lmm:SSL} to obtain
\begin{align}\lbeq{SSL2}
&\sum_{\substack{x,y\in\partial V_r\\ (x\ne y)}}\sum_{\substack{\partial\bn
 =\{o,g\}\\ \partial\bm=\vno}}\frac{w_{r,R}^h(\bn)}{Z_{r,R}^h}\,\frac{
 w_R(\bm)}{Z_R}\,\indic\{o\cn{\bn+\bm}{}x,y\text{ in }V_R\}\nn\\
&=\sum_{\substack{x,y\in\partial V_r\\ (x\ne y)}}\sum_{\substack{\partial\bn
 =\{x,g\}\\ \partial\bm=\{o,x\}}}\frac{w_{r,R}^h(\bn)}{Z_{r,R}^h}\,
 \frac{w_R(\bm)}{Z_R}\,\indic\{o\cn{\bn+\bm}{}y\text{ in }V_R\}.
\end{align}
For the event $o\cn{\bn+\bm}{}y$ in $V_R$ to occur under the source constraint
$\partial\bn=\{x,g\}$, $\partial\bm=\{o,x\}$, either one of the following
must be the case:
\begin{enumerate}[(i)]
\item
$o\cn{\bm}{}y$.
\item
$o\ncn{\bm}{}y$ and
$\exists u\in\overline{\cC_{\bm}(o)}\equiv\{v\in V_R:o\cn{\bm}{}v\}$
that is $(\bn+\bm')$-connected to $y$ in
$V_R\setminus\overline{\cC_{\bm}(o)}$,
where $\bm'$ is the restriction of $\bm$ on bonds
$b\subset V_R\setminus\overline{\cC_{\bm}(o)}$.
\end{enumerate}
Case (i) is easy; by Lemma~\ref{lmm:SSL}, the contribution to
\refeq{SSL2} is bounded as
\begin{align}\lbeq{error}
&\sum_{\substack{x,y\in\partial V_r\\ (x\ne y)}}\underbrace{\sum_{\partial\bn
 =\{x,g\}}\frac{w_{r,R}^h(\bn)}{Z_{r,R}^h}}_{=\Exp{\sigma_x}_{r,R}^h\le1}
 \sum_{\partial\bm=\{o,x\}}\frac{w_R(\bm)}{Z_R}\,\indic\{o\cn{\bm}{}y
 \}\nn\\
&\le\sum_{\substack{x,y\in\partial V_r\\ (x\ne y)}}
 \sum_{\substack{\partial\bm=\{o,x\}\\ \partial\bl=\vno}}\frac{w_R(\bm)}
 {Z_R}\,\frac{w_R(\bl)}{Z_R}\,\indic\{o\cn{\bm+\bl}{}y\}\nn\\
&=\sum_{\substack{x,y\in\partial V_r\\ (x\ne y)}}
 \underbrace{\sum_{\partial\bm=\{x,y\}}\frac{w_R(\bm)}{Z_R}}_{=
 \Exp{\sigma_x\sigma_y}_R}~\underbrace{\sum_{\partial\bl=\{o,y\}}\frac{w_R
 (\bl)}{Z_R}}_{=\Exp{\sigma_o\sigma_y}_R}.
\end{align}

\paragraph{\it Step 3: Conditioning on clusters.}
Case (ii) is a bit harder and needs extra care.  Here we use the
conditioning-on-clusters argument.  First, by conditioning on
$\overline{\cC_{\bm}(o)}$, we can rewrite the contribution
to \refeq{SSL2} from case (ii) as
\begin{align}\lbeq{conditioning}
\sum_{\substack{u\in V_R\\ x,y\in\partial V_r\\ (x\ne y)}}\sum_{\substack{A
 \subset V_R\\ (o,u,x\in A)}}\sum_{\substack{\partial\bn=\{x,g\}\\
 \partial\bm=\{o,x\}}}\frac{w_{r,R}^h(\bn)}{Z_{r,R}^h}\,\frac{w_R(\bm)}
 {Z_R}\,\indic\{\overline{\cC_{\bm}(o)}=A\}~\indic\{u\cn{\bn+\bm'}
 {}y\text{ in }V_R\setminus A\}.
\end{align}
Then, the sum over the current configurations in \refeq{conditioning} can be
rewritten as
\begin{align}
\sum_{\partial\bm=\{o,x\}}\frac{W_A(\bm)\,\cZ_{V_R\setminus A}}{Z_R}\,
 \indic\{\overline{\cC_{\bm}(o)}=A\}\sum_{\substack{\partial\bn=\{x,g\}\\
 \partial\bm'=\vno}}\frac{w_{r,R}^h(\bn)}{Z_{r,R}^h}\,\frac{W_{V_R
 \setminus A}(\bm')}{\cZ_{V_R\setminus A}}\,\indic\{u\cn{\bn+\bm'}
 {}y\text{ in }V_R\setminus A\}.
\end{align}
Now, by using Lemma~\ref{lmm:SSL}, the above expression is equal to
\begin{align}
\sum_{\partial\bm=\{o,x\}}\frac{W_A(\bm)\,\cZ_{V_R\setminus A}}{Z_R}\,
 \indic\{\overline{\cC_{\bm}(o)}=A\}\underbrace{\sum_{\partial\bn=\{u,x,y,g\}}
 \frac{w_{r,R}^h(\bn)}{Z_{r,R}^h}}_{=\Exp{\sigma_u\sigma_x\sigma_y}_{r,
 R}^h}~\underbrace{\sum_{\partial\bm'=\{u,y\}}\frac{W_{V_R\setminus A}(\bm')}
 {\cZ_{V_R\setminus A}}}_{=\Exp{\sigma_u\sigma_y}_{V_R\setminus A}},
\end{align}
where $\Exp{\sigma_u\sigma_y}_{V_R\setminus A}$ is the 2-spin expectation on
the vertex set $V_R\setminus A$ under the free-boundary condition, and is
bounded by $\Exp{\sigma_u\sigma_y}_R$ due to monotonicity.  As a result,
we obtain
\begin{align}\lbeq{leading}
\refeq{conditioning}&\le\sum_{\substack{u\in V_R\\ x,y\in\partial V_r\\
 (x\ne y)}}\Exp{\sigma_u\sigma_x\sigma_y}_{r,R}^h\,\Exp{\sigma_u\sigma_y}_R
 \sum_{\substack{A\subset V_R\\ (o,u,x\in A)}}\sum_{\partial\bm=\{o,x\}}
 \frac{W_A(\bm)\,\cZ_{V_R\setminus A}}{Z_R}\,\indic\{
 \overline{\cC_{\bm}(o)}=A\}\nn\\
&=\sum_{\substack{u\in V_R\\ x,y\in\partial V_r\\ (x\ne y)}}\Exp{\sigma_u
 \sigma_x\sigma_y}_{r,R}^h\,\Exp{\sigma_u\sigma_y}_R\sum_{\partial\bm=\{o,
 x\}}\frac{w_R(\bm)}{Z_R}\,\indic\{o\cn{\bm}{}u\}\nn\\
&\le\sum_{\substack{u\in V_R\\ x,y\in\partial V_r\\ (x\ne y)}}\Exp{\sigma_u
 \sigma_x\sigma_y}_{r,R}^h\,\Exp{\sigma_u\sigma_y}_R\sum_{\substack{\partial
 \bm=\{o,x\}\\ \partial\bl=\vno}}\frac{w_R(\bm)}{Z_R}\,\frac{w_R(\bl)}
 {Z_R}\,\indic\{o\cn{\bm+\bl}{}u\}\nn\\
&=\sum_{\substack{u\in V_R\\ x,y\in\partial V_r\\ (x\ne y)}}\Exp{\sigma_u
 \sigma_x\sigma_y}_{r,R}^h\,\Exp{\sigma_u\sigma_y}_R\underbrace{\sum_{\partial
 \bm=\{u,x\}}\frac{w_R(\bm)}{Z_R}}_{=\Exp{\sigma_u\sigma_x}_R}~
 \underbrace{\sum_{\partial\bl=\{o,u\}}\frac{w_R(\bl)}{Z_R}}_{=
 \Exp{\sigma_o\sigma_u}_R},
\end{align}
where, in the last line, we have used Lemma~\ref{lmm:SSL} again.

\paragraph{\it Step 4: Conclusion.}
Summarizing \refeq{1stbd}, \refeq{error} and \refeq{leading}, we arrive at
\begin{align}\lbeq{2ndbd}
\Exp{\sigma_o}_{r,R}^h\ge\frac{\dpst\bigg(\sum_{x\in\partial V_r}
 \Exp{\sigma_x}_{r,R}^h\,\Exp{\sigma_o\sigma_x}_R\bigg)^2}
 {\dpst\sum_{x,y\in\partial V_r}\Exp{\sigma_o\sigma_x}_R\,
 \Exp{\sigma_x\sigma_y}_R+\sum_{\substack{u\in V_R\\ x,y\in\partial V_r}}
 \Exp{\sigma_o\sigma_u}_R\,\Exp{\sigma_u\sigma_x}_R\,
 \Exp{\sigma_u\sigma_y}_R\,\Exp{\sigma_u\sigma_x\sigma_y}_{r,R}^h}.
\end{align}
Now we take $h\uparrow\infty$ in both sides.  In this limit, the spins on
$\partial V_r$ take on $+1$.  Moreover, by Griffiths' inequality, we have
$\lim_{h\uparrow\infty}\Exp{\sigma_u\sigma_x\sigma_y}_{r,R}^h=
\Exp{\sigma_u}_{r,R}^\infty\le\Exp{\sigma_o}_{\mathrm{dist}(u,\partial V_r)}^+$.
Therefore,
\begin{align}
\Exp{\sigma_o}_r^+\ge\frac{\dpst\bigg(\sum_{x\in\partial V_r}
 \Exp{\sigma_o\sigma_x}_R\bigg)^2}
 {\dpst\sum_{x,y\in\partial V_r}\Exp{\sigma_o\sigma_x}_R\,
 \Exp{\sigma_x\sigma_y}_R+\sum_{\substack{u\in V_R\\ x,y\in\partial V_r}}
 \Exp{\sigma_o\sigma_u}_R\,\Exp{\sigma_u\sigma_x}_R\,
 \Exp{\sigma_u\sigma_y}_R\,\Exp{\sigma_o}_{\mathrm{dist}(u,\partial V_r)}^+}.
\end{align}
Taking $R\uparrow\infty$, we finally obtain \refeq{correlation}.
\QED

\subsection{Proof of the main theorem}\label{ss:proof}
\proof[Proof of Theorem \ref{thm:main}.]
We proceed indirectly and assume, by contradiction, that \refeq{main} is
false.  Then there exist a constant $K>0$ and a monotone sequence 
$(r_k)_{k\in\mathbb N}$ diverging to $\infty$ such that
\begin{align}\lbeq{lowerBd2}
\Exp{\sigma_o}_{r_k}^+ \le K r_k^{-(1+\vep)}
\end{align}
whenever $k$ is large enough.

We are starting from Proposition \ref{prp:2ndmom}.  We estimate every term in
the numerator and the denominator of \refeq{correlation} using \refeq{twopt}.
Firstly, the numerator of \refeq{correlation} is of the order $r^2$ since
\begin{align}
\sum_{x\in\partial V_r}\Exp{\sigma_o\sigma_x}\asymp\sum_{x\in\partial V_r}
 |x|^{2-d}\asymp r^{d-1}r^{2-d}=r.
\end{align}
Secondly, the first term in the denominator is of order $O(r^{2})$ since
\begin{align}
\sum_{x,y\in\partial V_r}\Exp{\sigma_o\sigma_x}\Exp{\sigma_x\sigma_y}\asymp
 \sum_{x\in\partial V_r}\veee{x}^{2-d}\sum_{y\in\partial V_r}\veee{x-y}^{2-d}
 \asymp r^{d-1}r^{2-d}r = r^2,
\end{align}
where we have used $\veee{\cdot}=|\cdot|\vee1$ (cf. below
\refeq{perc-correlation}).  The second term in the
denominator is the dominant one.  To this end, we fix a sequence $r_k$
satisfying \refeq{lowerBd2}.
We split
the sum over $u$ into three cases: (i) $|u|<{r_k}/2$,
(ii) ${r_k}/2\le|u|<3{r_k}/2$, (iii) $3{r_k}/2\le|u|$,
and show that it is $O({r_k}^3)$ for any $\vep>0$.\\
Case (i):
\begin{align}
&\sum_{\substack{|u|<{r_k}/2\\ x,y\in\partial V_{r_k}}}\veee{u}^{2-d}
 \veee{u-x}^{2-d}\veee{u-y}^{2-d}\Veee{r_k-|u|}^{-(1+\vep)}\nn\\
&\quad\asymp r_k^{2(2-d)-(1+\vep)}\underbrace{\sum_{x,y\in\partial
 V_{r_k}}\sum_{|u|<{r_k}/2}\veee{u}^{2-d}}_{\asymp r_k^{2(d-1)+2}}\nn\\
&\quad\asymp r_k^{3-\vep}.
\end{align}
Case (ii):
\begin{align}
&\sum_{\substack{r_k/2\le|u|<3r_k/2\\ x,y\in\partial V_{r_k}}}|u|^{2-d}
 \veee{u-x}^{2-d}\veee{u-y}^{2-d}\Veee{r_k-|u|}^{-(1+\vep)}\nn\\
&\quad\asymp r_k^{2-d}\sum_{r_k/2\le|u|<3r_k/2}\Veee{r_k-|u|}^{-(1+\vep)}
 \underbrace{\sum_{x\in\partial V_{r_k}}\veee{u-x}^{2-d}\sum_{y\in\partial
 V_{r_k}}\veee{u-y}^{2-d}}_{\asymp r_k^2}\nn\\
&\quad\asymp r_k^{4-d}\int_{r_k/2}^{3r_k/2}\veee{r_k-l}^{-(1+\vep)}l^{d-1}
 dl\nn\\
&\quad\asymp r_k^3\int_0^{{r_k}/2}\veee{l}^{-(1+\vep)}dl~\asymp{r_k}^3.
\end{align}
Case (iii): By the Schwarz inequality,
\begin{align}
&\sum_{\substack{|u|\ge3r_k/2\\ x,y\in\partial V_{r_k}}}|u|^{2-d}|u-x|^{2-d}
 |u-y|^{2-d}(|u|-{r_k})^{-(1+\vep)}\nn\\
&\quad\asymp r_k^{2-d-(1+\vep)}\underbrace{\sum_{\substack{x,y\in\partial
 V_{r_k}}}\sqrt{\sum_{|u|\ge3r_k/2}|u-x|^{4-2d}}\sqrt{\sum_{|u|\ge3r_k/2}
 |u-y|^{4-2d}}}_{\asymp r_k^{2(d-1)+4-d}}\nn\\
&\quad\asymp r_k^{3-\vep}.
\end{align}

Plugging these estimates into the bound of Proposition \ref{prp:2ndmom} obtains
\begin{equation}\lbeq{lowerbd}
	\Exp{\sigma_o}_{r_k}^+\ge C\frac{{r_k}^2}{{r_k}^2+{r_k}^3}\asymp r_k^{-1}
\end{equation}
for any large $k$ and for some $C$ (independent of $k$), which contradicts 
\refeq{lowerBd2}, and \refeq{main} follows.

The claim $\rho\le1$ follows straightforwardly: Suppose $\rho>1+\vep$ for some $\vep>0$, then there exists a sequence $(r_k)_{k\in\mathbb N}$ such that
\[\frac{\log {\Exp{\sigma_o}_{r_k}^+}}{\log {r_k}}<-1-\vep,\]
and this contradicts \refeq{main}.
\QED

\section*{Acknowledgements}
The work of AS is supported by the JSPS Grant-in-Aid for Challenging
Exploratory Research  15K13440.
The work of SH is supported by the Ministry of Education, Culture, Sports,
Science and Technology through Program for Leading Graduate Schools
(Hokkaido University ``Ambitious Leader's Program'').
We thank Aernout van Enter for providing references about the history of
the problem.

\end{document}